\documentclass[letter]{aa}  

\usepackage{graphicx}
\usepackage{txfonts}
\usepackage{natbib, hyperref}
\usepackage{float}
\usepackage{soul}
\usepackage{xcolor}

\usepackage{lineno}
\linenumbers

\renewcommand\linenumberfont{\color{white}}
\begin{document} 

   \title{Not Just Gas: How Solid-Driven Torques Shaped the Migration of the Galilean Moons}

   \subtitle{}

%%%%%%%%%%%%%%%%%%%%%%%%%%%%%%%%%%%%%%%%
% Please do not include ORCIDs next to author names.
% Only ORCIDs authenticated by individual authors in EDP Sciences editorial system will be taken into account.
% ORCIDs included here will be removed.
%%%%%%%%%%%%%%%%%%%%%%%%%%%%%%%%%%%%%%%%

   \author{L. Gonzalez-Rivas\inst{1}, L. Krapp\inst{1}\inst{2}, X. Ramos\inst{3}, P. Benitez-Llambay\inst{4}, 
        }
        
    \institute{Departamento de Astronomía, Facultad de Ciencias Físicas y Matemáticas Universidad de Concepción, Av. Esteban Iturra s/n Barrio Universitario, Casilla 160-C, Chile\\
              \email{lkrapp@udec.cl}
         \and
             Department of Astronomy and Steward Observatory,  University of Arizona, Tucson, Arizona 85721, USA 
             \and
    Centro Multidisciplinario de Física, Vicerrectoría de Investigación, Universidad Mayor, 8580745 Santiago, Chile
\and
Facultad de Ingenier{\'\i}a y Ciencias Universidad Adolfo Ib\'a\~nez, Av. Diagonal las Torres 2640, Pe\~nalol\'en, Chile}

   \date{}

% \abstract{}{}{}{}{}
% 5 {} token are mandatory
 
  \abstract
   {Surviving rapid inward orbital migration is a crucial aspect of formation models for the Jupiter's Galilean moons. }
   {The primary aim of this study is to investigate the orbital migration of the Galilean moons by incorporating self-consistent solid dynamics in circumjovian disk models.} 
   {We perform two-fluid simulations using the FARGO3D code on a 2D polar grid. The simulations model a satellite with the mass of a proto-moon, Europa, or Ganymede interacting with a circumjovian disk. The dust component, coupled to the gas via a drag force, is characterized by the dust-to-gas mass ratio ($\epsilon$) and the   Stokes number  ($T_s$).}
   { The effect of solids fundamentally alter the satellites' evolution. We identify a vast parameter space where migration is slowed, halted, robustly reversed -leading to outward migration-, or significantly accelerated inward.
   The migration rate is dependent on satellite mass, providing a natural source of differential migration.}
   {Solid dynamics provides a robust and self-consistent mechanism that fundamentally alters the migration of the Galilean moons, potentially addressing the long-standing migration catastrophe. This mechanism critically affects the survival of satellites and could offer a viable physical process to explain the establishment of resonances through differential migration. These findings establish that solid torques are a critical, non-negligible factor in shaping the final architecture of satellite systems.}
   

\titlerunning{Solid-driven migration of Galilean Moons}
   \maketitle
%
%-------------------------------------------------------------------

\section{Introduction}

The formation of the Galilean satellites (Io, Europa, Ganymede, and Callisto) is a cornerstone problem in planetary theory, offering a scaled-down analog of planetary system formation. 
Paralleling this process, the moons are believed to have formed via bottom-up accretion process within Jupiter's solid-rich primordial circumplanetary disk \citep{Lunine,Coradini1995, Peale}, analogous to the initial phase planet formation \citep{Pollack1996}.

While foundational models such as \cite{Canup2002} and \cite{Mosqueira2003a} successfully predict the system’s total mass and compositional gradient, these frameworks have long been crippled by the Type-I migration catastrophe. This is the tendency for gravitational torques from the massive gaseous phase of the circumjovian disk to drive proto-moons rapidly inward \citep{Goldreich1980, Ward1997}, resulting in their destruction \citep{Mosqueira2003b, Canup2006, Sasaki2010,Ogihara2012, Miguel2016}.
Thus, moon survival hinges on counteracting this rapid inward migration. 
%

%%%%%%%%%%%%%%%%%%%%%%%%%%%%%%%%%%%%%%%%%%%%%%%%%%%%%%%%%%%%%%
\begin{figure*}[t!]
   \centering
   \includegraphics[scale=0.7]{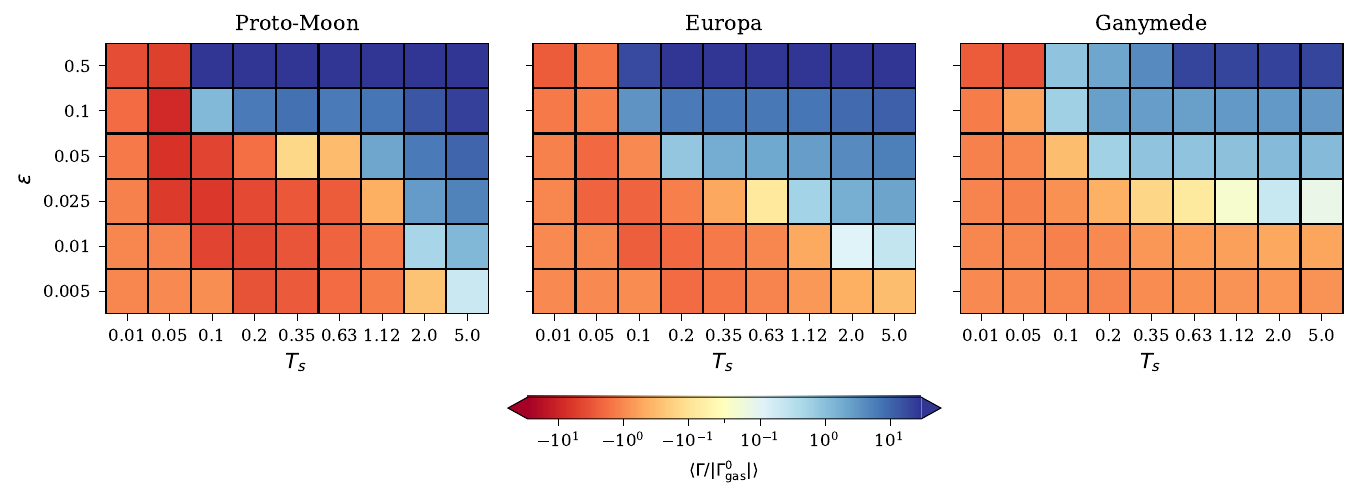}
      \caption{Normalized time-averaged net torque for all the models  considered in this work, as a function of initial dust-to-gas density ratio and Stokes number. 
      From left to right, the mass of the satellite corresponds to $M_s=1.25\times 10^{-5} M_J$, $M_s=2.53\times 10^{-5} M_J$, and $M_s=7.82\times 10^{-5} M_J$, respectively.
      Blue color indicates a net positive torque (outward orbital migration). 
      The normalization, $\Gamma^0_{\rm gas}$, corresponds to the net torque when $\epsilon=0$. From left to right,  $\Gamma^0_{\rm gas} = -3.63 \Gamma_0$, $\Gamma^0_{\rm gas} = -3.33\Gamma_0$ and $\Gamma^0_{\rm gas} = -2.83 \Gamma_0$, with $\Gamma_0 = (M_s/M_J)^2 \Sigma_0 h^{-2} r^4_0 \Omega^2_0$. }
         \label{fig1}
         \vspace{-2.5ex}
   \end{figure*}
%%%%%%%%%%%%%%%%%%%%%%%%%%%%%%%%%%%%%%%%%%%%%%%%%%%%%%%%%%%%%%
A promising mechanism that can mitigate the inward migration naturally emerges from the aerodynamics of solids \citep{Benitez-Llambay2018}. 
The scattering of solids by the satellite's gravitational potential can drive large asymmetries in the flow, critically affecting the net torques exerted on the satellite.
However, the implications of this migration regulated by solids have been addressed solely in the context of planetary formation \citep[e.g.,][]{Regaly2020, Chrenko2024, Hou024}.

In this Letter, we present the first systematic investigation of the migration of the Galilean moons incorporating self-consistent solid dynamics. We demonstrate that torques arising from solid dynamics fundamentally alter the net torque on the satellites, causing it to deviate significantly from gas-only models. This effect is decisive for the migration history of the Galilean moons, as it can slow, accelerate, or even reverse the migration depending on the local dust-to-gas ratio and particle Stokes number.
We summarize our findings in Figure\,\ref{fig1}, showcasing the solid's torque relative to the gaseous torque, exerted by the circumjovian disk on satellites with the mass of Europa and Ganymede. 
%

%%%%%%%%%%%%%%%%%%%%%%%%%%%%%%%%%%%%%%%%%%%%%%%%%%%%%%%%%%%%%%

%
%
\vspace{-3ex}
\section{Methods}

\noindent We perform two-fluid simulations with the FARGO3D code \citep{Benitez-Llambay2016, Benitez-Llambay2019} on a 2D polar grid $(r, \phi)$ with $768 \times 4096$ cells spanning $r \in [0.5 r_0, 1.5 r_0]$, which is enough to account for the main source of gaseous torque.
A satellite of mass $M_{\rm s}$ is on a circular orbit around a central planet\footnote{We adopt a planetocentric frame and include the indirect terms arising from the planet and the disk (dust+gas) \citep{Crida2025}.}  of mass $M_{\rm J}$, with semi-major axis $r_s$ and orbital frequency $\Omega_s$. We adopt units such that our reference time, mass and distance units are $\Omega^{-1}_0=1$, $M_0=1$, $r_0=1$, respectively. We set $M_J=M_0$, $r_s=r_0$, which implies $\Omega_s=\Omega_0$. The size of each cell in the hydrodynamical grid is $\sim 1.3\times 10^{-3} r_0$, which is also enough to resolve the small-scale substructure induced by the moon's gravitational perturbation on the solid phase of the disk.
The satellite's potential is modeled as a Plummer potential with a smoothing length equal to the moon's Hill radius. Initial and boundary conditions are taken from \citep{Benitez-Llambay2018}. 
The gaseous phase of the disk is locally isothermal with an aspect ratio $h=0.1$ \citep{Brunton2025} and viscosity $\nu = \alpha c_{\rm s} h r$ with $\alpha = 10^{-3}$. Its initial surface density is $\Sigma(r) = \Sigma_0 (r/r_0)^{-1/2}$, where $\Sigma_0 = 6 \times 10^{-4}\, M_{\rm J}/r_0^2$. The locally isothermal assumption is adopted for computational feasibility, but we acknowledge that it neglects complex thermodynamic effects such as cold thermal torques and heating torques, which can affect the gaseous torque significantly \citep{Lega2014, Benitez-Llambay2015, Masset2017}.
The dust component is initialized with a constant dust-to-gas mass ratio, $\epsilon$, and is coupled to the gas via a drag force consistent with the Epstein and Stokes regimes. \citep{Weidenschilling1977}. Dust back-reaction onto the gas is enabled by default.
In our exploration, we consider nine Stokes numbers\footnote{We define the Stokes number as the dimensionless stopping time, expressed in units of the local Keplerian frequency of the disk.} $T_s = [0.01,0.05,0.1,0.2,0.35,0.63,1.12,2,5]$ and seven different dust-to-gas mass ratios $\epsilon = [0, 0.005, 0.01, 0.025, 0.05, 0.1, 0.5]$, consistent with subsolar to supersolar metallicity values.
For all these parameters, we consider three different moon's masses, which are consistent with a proto-moon ($M_{\rm s} =1.25\times10^{-5} \,M_{\rm J} $), Europa ($M_{\rm s} =2.53\times10^{-5} \, M_{\rm J} $), and Ganymede ($M_{\rm s} =7.82\times10^{-5} \, M_{\rm J}$). In this paper we focus on the late stages of moon formation, therefore we do not explore satellite masses below $M_s \lesssim 1.25 \times 10^{-5} M_J$. Nevertheless, based on results from previous work, dust torques are positive and dominant for low-mass bodies, consistent with the ones not considered in our exploration \citep{Benitez-Llambay2018}.
We integrate the system until the torque reaches steady-state, which varies varies between 28 to 56 satellite orbits, depending on the parameters.

\vspace{-3ex}
\section{Torque maps for Jovian disks}
\label{sec:results}

We present the main results of our numerical exploration in Figure~\ref{fig1}. Each panel summarizes our results for a different satellite mass. The colors represent the net (gas+dust) torque exerted by the circumjovian disk, normalized by the torque measured in a simulation in which the solid phase is not taken into account. 
The total torque includes contributions from the disk potential and the indirect potential arising from the non-inertial frame (see e.g. Eqs. 5–7 \cite{Crida2025}). 
While in most cases the torque reaches a steady-state value within the integration time, for large dust-to-gas mass ratios the torque is not strictly steady. For this reason, we characterize the torque by its time-averaged value over the final orbital period of the integration. 
In what follows, we present our main results.
\vspace{-3ex}
\paragraph{Positive torques are ubiquitous:} A first striking result of our exploration is that solids can induce positive torques (hence outward migration) for all satellite masses considered, provided the Stokes number ($T_s$) and dust-to-gas mass ratio ($\epsilon$) exceed certain thresholds. 
This dependency is clearest for the proto-Moon and Europa. 
For these bodies, positive torque at low solid fractions (e.g., $\epsilon < 0.01$) requires large Stokes numbers ($T_s \gtrsim 2$). As the solid fraction $\epsilon$ increases, this requirement on $T_s$ relaxes, and models with progressively smaller Stokes numbers also yield positive torques. 
The proto-Moon's behavior closely mirrors Europa's, though its transition to outward migration occurs at slightly higher Stokes numbers. 
For the more massive Ganymede, the conditions are slightly more restrictive as outward migration is possible only when $\epsilon \gtrsim 0.025$ and $T_s \gtrsim 1$.
\vspace{-5.5ex}
\paragraph{Torques differ from those in gas-only disks:} In most cases explored in this work, the net torque is affected by the presence of dust whenever $T_s > 0.01$. Critically, we find a dramatic enhancement of negative torques for the proto-Moon. This is a key result: within the intermediate Stokes number regime ($0.1 \lesssim T_s \lesssim 1$), we find a net torque magnitude $|\Gamma| \sim  3 -7 \times |\Gamma^0_g|$, even at subsolar metallicities. Here $\Gamma^0_g$ corresponds to the net torque in our dust-free models.  
While enhanced negative torques can accelerate orbital loss for proto-moons, this effect depends on the local Stokes number and solid fraction. If accretion allows the moon to grow faster than it migrates, it can bypass this unstable regime to reach larger masses (like Europa or Ganymede), where outward migration and orbital stalling become possible.
A second striking feature, shown in Figure~\ref{fig1}, is the large positive torque achieved at high metallicity ($\epsilon \sim 0.5$). This torque is over two orders of magnitude larger than in a gas-only model. We found this regime to be strongly influenced by dust feedback, as vortices forming in the horseshoe region induce stochastic fluctuations in the dust torque, consistent with findings by \cite{Chen2018} in disks with very large dust-to-gas mass ratios. 

\vspace{-3ex}

\section{Discussion}
\label{sec:discusion}

The dust torque exerts a significant influence on satellite migration, either promoting outward migration or critically accelerating inward migration within specific parameter regimes. 
The division between these two migration behaviors is fundamentally determined by the satellite mass, the particle Stokes number, and the solid mass fraction.
A general trend, visually represented in Figure~\ref{fig1}, is that outward migration typically occurs for large stokes numbers and moderate to large gas-to-dust mass ratios.
Crucially, since our torque measurements are done on a fixed planetary orbit, the sign of the torques are independent of the disk surface density, $\Sigma$.
Therefore, the migration regimes highlighted in Figure~\ref{fig1} are applicable across a wide range of formation scenarios, including gas-starved disk models \citep{Canup2002} and minimum mass subnebula models \citep{Lunine, Mosqueira2003a} as long as the migration rate remains small compared to the radial drift of solids.
In the following we discuss potential implications or modifications of our findings under more realistic physical modeling.
\vspace{-3ex}
\subsection{Implications for Circumjovian Disk Models}

In gas-starved disk models, the solid-to-gas ratio has been suggested to increase significantly, potentially approaching unity \citep{Canup2002}. Our results indicate that if $T_s > 0.1$, such a scenario would robustly lead to an outward migration regime of the regular sattelites as they grow beyond $M_s \sim 10^{-5} M_J$. 
Given that typical average surface densities in these models ($\Sigma \sim 10 - 1000 \, \text{g}/\text{cm}^2$) are consistent with $T_s \gtrsim 0.1$ for solids in the centimeter-to-meter size range, this outward migration is a highly plausible outcome.
Therefore, solid driven migration directly impacts the orbital evolution and final formation locations of the Galilean moons. 

For models with augmented solid fractions, but still well below unity \citep[e.g., Solid Enhanced Minimum Mass disk][]{Estrada2009}, outward migration is triggered as the Stokes number exceeds unity ($T_s > 1$). However, in this models the average gas surface densities can exceed $\Sigma \sim 1000 \, \text{g}/\text{cm}^2$, and therefore solids larger than a meter are required to reach $T_s \sim 1$, until gas starts dissipating. Thus, if the particle size distribution in the primordial circumjovian disk is limited below the "meter barrier," the solids are instead likely to drive fast inward migration.
Furthermore, our results are consistent with rapid inward migration under subsolar metallicity conditions, such as the torques shown for $\epsilon = 0.005$ in Figure~\ref{fig1} only for the proto-moon and Europa. This finding suggests that migration rates in existing formation models based purely on pebble accretion may be significantly underestimated\citep{Shibaike2019,Ronnet2020,Madeira2021}. Overall, the complex and non-trivial dependence of the net torque on $\epsilon$ and $T_s$ necessitates that modern formation models \citep[e.g., ][]{Shibaike2025} must now incorporate the time-evolving solid component as a critical source of migration to accurately track satellite orbital evolution.

The possibility of outward migration or stalling naturally raises the question of mass regulation. If solid driven torques prevent orbital decay, we propose that mass regulation is dictated by the solid reservoir or the dissipative timescale of the circumplanetary disk.
In this scenario, growth concludes when the available solids in the feeding zone are exhausted or when the gas disk (which mediates these torques) dissipates. This shifts the focus from a continuous flux model to a reservoir-limited scenario, where the final system architecture is primarily determined by the total solid budget.
\vspace{-3ex}
\subsection{Differential Migration of Galilean Moons} 
\vspace{-2ex}
The differing migration regimes found for Ganymede and Europa are a strong indication that solids provide a viable source of differential migration, a crucial requirement for locking the regular satellites into their observed mean motion resonances. 
In the case of Ganymede, outward migration occurs when the conditions are $T_s \gtrsim 0.1$ and $\epsilon \gtrsim 0.02$. Although Ganymede is typically prone to the fastest inward migration in gas-only models, this rapid decline is efficiently mitigated if it reached its final mass within a region characterized by a solid fraction augmented by a few percent.
On the other hand, Europa shows a transition to outward migration occurring even for lower dust-to-gas ratios of $\epsilon \sim 0.005$, provided the Stokes number is high enough ($T_s \gtrsim 5$).  
This suggests that Europa may require local disk conditions with larger particle sizes and/or higher solid-to-gas densities than Ganymede to reach an outward migration regime.
Although Figure\,\ref{fig1} uses a mass-dependent normalization, the transition between inward and outward migration varies for Europa and Ganymede across the $(\epsilon, T_s)$ parameter space. This mass-dependent response enables the differential migration required to reach mean-motion resonances. For instance, in a fixed disk model, a more massive Ganymede may experience reduced inward migration while a smaller Europa stalls or moves outward, facilitating orbital convergence.
While detailed simulations for Io and Callisto are beyond the scope of this Letter, their intermediate masses suggest they would experience comparable, intermediate trends, further supporting the general applicability of solid-driven migration across the entire system.
\vspace{-3ex}
\subsection{The Role of Solid Accretion} 
\vspace{-1.5ex}
\label{sec:accretion}
While our work calculates steady-state torques on fixed non-accreting satellites, the formation of the Galilean moons relies heavily on accretion of solids  \citep[see e.g.,][]{Ormel2024}, a process that fundamentally alters the total torque in two ways. 
First, accretion modifies the spatial distribution of solids by removing material close to the moon, which tends to accentuate the asymmetry in the solid distribution \citep{Regaly2020, Chrenko2024}. This strengthening effect means that the magnitude of the torque exerted by solids can increase with accretion strength. For satellite masses smaller than the ones considered in this work, the torque exerted by solids can significantly exceed the gas torque due to accretion \citep{Chrenko2024}, or for larger masses provided the metallicity is large enough as we showed in our work.
Second, accretion introduces the so-called accretion torque, which is produced by the angular momentum carried by the accreted solids \citep{Chrenko2024}. This torque can be a substantial component that can either diminish the net torque by solids or strengthen it, depending on the Stokes number. Studies incorporating self-consistent accretion in the context of planet formation demonstrated that the positive magnitude of the dust torque is notably larger for low-mass bodies and small Stokes numbers when accretion is included \citep{Guilera2025}. 
These findings strengthen the conclusion that outward migration can be a natural outcome for moons forming via solid accretion, not only at late stages, but more importantly during its formation stage. 

\vspace{-3ex}
\subsection{Thermal torques} 
\label{sec:thermodyanmics}

Our analysis relies on a locally isothermal disk model, simplifying the computation of gas torques. However, realistic disk thermodynamics introduces critical non-isothermal effects that interact directly with the dust dynamics and exert additional and substantial torques.
One such effect is the so-called cold thermal torque \citep{Lega2014}, which arises from thermal diffusion close to the moon. However, the thermal structure is inherently connected to the dust component: dust feedback can significantly reshape the cold, dense lobes around the planet, which are the source of the thermal torque \citep{Chametla2025}. 
Furthermore, the release of energy during accretion generates a heating torque \citep{Benitez-Llambay2015}.
The strong asymmetries in the dust distribution around the planet could alter the local opacity, potentially leading to an asymmetric thermal diffusivity near the moons. 
Such an effect could affect the net thermal torque acting on the satellite. Therefore, a comprehensive picture of Galilean moon migration requires integrating these non-isothermal physics, as the gravitational torque from solids and thermal torques operate under coupled conditions.
\vspace{-3ex}
\subsection{Caveats and Future Directions}

Our primary finding is that torques arising from solid dynamics cannot be neglected in understanding the migration history of the Galilean moons.
While simplified, our 2D framework facilitates broad and systematic parameter exploration with self-consistent solid dynamics. 
However,  two-dimensional simulations may overestimate dust-feedback torques compared to three-dimensional disks at the same dust-to-gas ratio \citep{Hsieh2020}. Moreover, disk asymmetries and filaments are sensitive to turbulent diffusion and vertical settling, which likely impact the net torque, and may lead to non-steady migration \citep{Chametla2025}. 
In addition, 3D models are required to validate the effect of the gravitational softening, which can artificially enhance gaseous torque while suppressing dusty torque, although recent work suggests a weak sensitivity for low Stokes numbers \citep{Chametla2025}.
Finally, our torque values assume a fixed circular orbit and neglect mass growth. Realistic evolutionary tracks demand coupled migration and accretion simulations that include the accretion torque \citep{Chrenko2024}, which can substantially alter the total torque. Migration itself modifies relative dust–moon velocities, further affecting torque asymmetry.
Due to the low eccentricity of Galilean moons, we assumed circular orbits. However, thermal forces and dust feedback can excite eccentricity \citep{Eklund2017}, which strongly influences asymmetry morphology and torque magnitude. Future models must solve the non-isothermal energy equation to capture eccentricity growth and its interplay with dust-feedback torques self-consistently.
\vspace{-3ex}
\section{Conclusions}

We presented the first systematic investigation into the influence of solid dynamics on the orbital evolution of Jupiter’s Galilean moons, suggesting that torques arising from the solid component are a critical element that must be integrated into future satellite formation models to achieve a complete picture of their evolution. 
Our results demonstrate that the total torque experienced by forming satellites deviates significantly from gas-only models across a wide range of parameters. 
These findings elucidate a self-consistent migration mechanism that aligns seamlessly with classical and contemporary Jovian disk models assuming elevated dust-to-gas mass fractions \citep[e.g.,][]{Cilibrasi2018, Batygin2020, Schneeberger2025}. 
Crucially, the differing migration regimes found for the moons provide a viable source of differential migration.  A successful model must identify parameters that not only reproduce the differential migration needed to capture Io, Europa, and Ganymede in Laplace resonances \citep[][]{Peale2002} but also simultaneously explain the moons' compositional gradient \citep[][]{Kuskov2005,Canup2009}. 
Finally, while this work focuses on Jovian moons, solid-driven migration could be pivotal in shaping exomoon systems \citep{Heller2014, Kipping2022} and other regular satellites in our solar system \citep{Coradini1995, Alibert2007}. These insights underscore that coupling the time-evolving solid component with gas dynamics, a key aspect of circumplanetary disk models \citep{Krapp2024}, is mandatory for accurately tracking satellite orbital evolution and developing a comprehensive understanding of satellite system architectures across diverse planetary environments.

%}
 
%%%%%%%%%%%%%%%%%%%%%%%%%%%%%%%%%%%%%%%%%%%%%%%%%%%%%%%%%%%%%%
\begin{acknowledgements}
L. K. thanks Robin Canup, Yuri Fujii, and Konstantin Batygin
for inspiring discussion that led to this publication during the Circumplanetary Disks and Satellite Formation III conference in Kyoto, Japan.
L. K. acknowledges support from
the Heising-Simons Foundation, ANID Fondecyt Iniciacion (project 11250447), ANID-Quimal 220002, and ANID BASAL project FB21003. L. K. acknowledges High Performance Computing resources supported by the University of Arizona TRIF, UITS, and Research, Innovation, and Impact (RII) and maintained by the UArizona Research Technologies department.
P.~B.~L. acknowledges  support from ANID, QUIMAL fund ASTRO21-0039 and FONDECYT project 1231205.
\end{acknowledgements}
\vspace{-6.5ex}
%

\bibliographystyle{aa} % style aa.bst
\bibliography{bib} % your references Yourfile.bib

\begin{appendix}
\section{Dust Density and Torque Contributions}

The contribution of the solid material to the net torque is driven by an asymmetric density distribution in the horseshoe region. This asymmetry results from the satellite scattering solids as they drift through the disk.
As a result of this scattering, a complex structure develops. For large Stokes Numbers this typically resembles an under-dense cavity with a lower density in the region trailing the satellite, which is itself asymmetric, and an over-dense stream of dust leading the satellite's orbit. % spiraling towards the central planet.
The outcome of these complex asymmetric flows is generally a positive torque. However, this is not always the case; as shown in Figure~\ref{fig1}, solids can also increase the net negative torque~\citep{Benitez-Llambay2018}.

\begin{figure}[]
   \centering
   \includegraphics[scale=0.95]{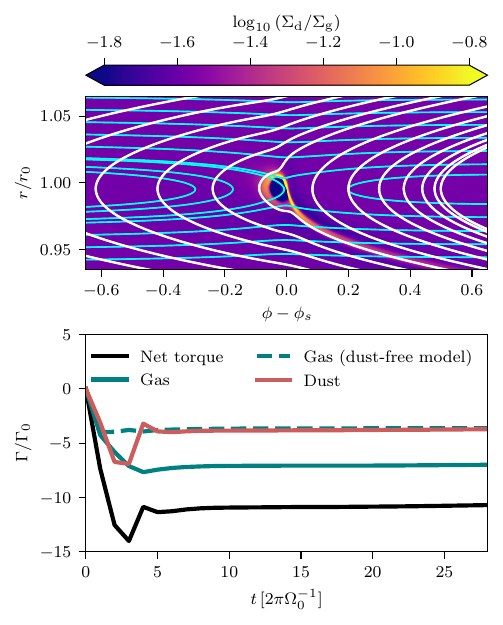}
      \caption{Top row: Dust surface density (color) with superimposed gas (cyan) and dust (white) velocity streamlines. Bottom row: Time evolution of the normalized torques exerted on the embedded moon. The solid black curve shows the net (gas + dust) torque, while the red and green curves correspond to the dust and gas contributions, respectively. The dashed green curve indicates the net gas torque in the corresponding dust-free reference simulation. Simulation corresponds to Proto-Moon with  $\epsilon = 0.025$ and $T_s = 0.35$. }
    \label{fig2}
   \end{figure}

\begin{figure}[]
   \centering
   \includegraphics[scale=0.95]{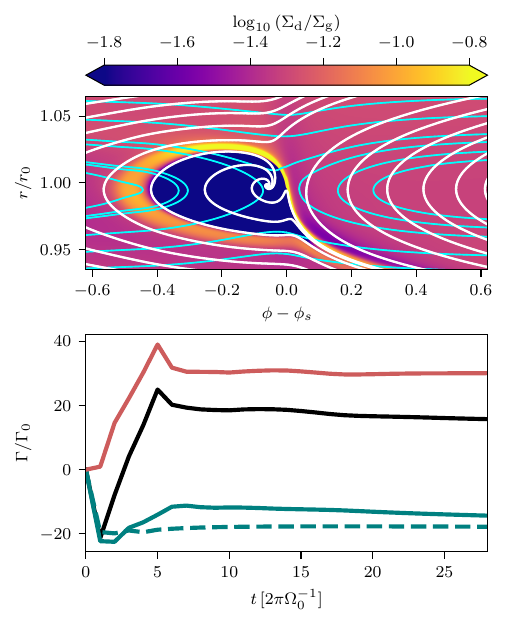}
      \caption{Same as Figure~\ref{fig2} but for Ganymede with $\epsilon = 0.05$ and $T_s = 0.35$. }
    \label{fig3}
   \end{figure}

We illustrate this density structure in Figures~\ref{fig2} (Proto-Moon) and~\ref{fig3} (Ganymede). %, which shows the dust-to-gas density ratio. % for the Proto-moon and Ganymede cases. 
The overlaid gas and dust streamlines highlight the key dynamical processes responsible for this distribution: the radial drift of dust particles and the asymmetric flows in the co-orbital region.
In the Proto-moon case, only a small dust cavity forms. As a result, the dust torque is dominated by the radially drifting dust stream interior to the orbit, yielding a strongly negative dust torque contribution and thus a net negative torque on the satellite.
In contrast, the more massive Ganymede forms a much larger dust cavity. This allows the positive dust torque to dominate, whose absolute value is larger than the gas torque (depending on the dust-to-gas mass fraction), resulting in a net positive torque.
The bottom panels of Figures~\ref{fig2} and~\ref{fig3} show the time evolution of the total torque for both cases. For reference, we also include the torque obtained from dust-free simulations. In both the Protomoon and Ganymede cases, the presence of dust substantially modifies the gas torque through dust and gas backreaction, underscoring the critical importance of dust feedback in these models.

%%%%%%%%%%%%%%%%%%%%%%%%%%%%%%%%%%%%%%%%%%%%%%%%%%%%%%%%%%%%%%
%%%%%%%%%%%%%%%%%%%%%%%%%%%%%%%%%%%%%%%%%%%%%%%%%%%%%%%%%%%%%%
%%%%%%%%%%%%%%%%%%%%%%%%%%%%%%%%%%%%%%%%%%%%%%%%%%%%%%%%%%%%%%
%%%%%%%%%%%%%%%%%%%%%%%%%%%%%%%%%%%%%%%%%%%%%%%%%%%%%%%%%%%%%%

\end{appendix}

\end{document}